\documentclass[11pt]{article}

\usepackage[latin1]{inputenc}
\usepackage[T1]{fontenc}
\usepackage{mathpazo}
\usepackage{geometry}
\usepackage{amsmath}
\usepackage{graphicx}
\usepackage{float}
\usepackage{authblk}
\usepackage{bm,upgreek}
\usepackage[labelfont=bf]{caption}

\ifpdf
\usepackage[hyperindex=true,
colorlinks=true,
hidelinks]{hyperref}
\else
\usepackage[hyperindex=true,
colorlinks=false,
hidelinks]{hyperref}
\fi

\usepackage[all]{hypcap}

\title{Aberration-free calibration for 3D single molecule localization microscopy}

\author[1,*]{Clément Cabriel}
\author[1]{Nicolas Bourg}
\author[2]{Guillaume Dupuis}
\author[1,**]{Sandrine Lévêque-Fort}
\affil[1]{Institut des Sciences Moléculaires d'Orsay (ISMO), CNRS UMR 8214, Université Paris-Sud, bâtiment 520, rue André Rivière, 91405 Orsay Cedex, France}
\affil[2]{Centre de Photonique BioMédicale (CPBM), CNRS FR 2764, Fédération LUMAT, Université Paris-Sud, bâtiment 520, rue André Rivière, 91405 Orsay Cedex, France}
\affil[*]{Corresponding author: clement.cabriel@u-psud.fr}
\affil[**]{Corresponding author: sandrine.leveque-fort@u-psud.fr}

\date{\today}

\begin{document}

\maketitle

{\bfseries We propose a straightforward sample-based technique to calibrate the axial detection in 3D single molecule localization microscopy (SMLM). Using microspheres coated with fluorescent molecules, the calibration curves of PSF-shaping- or intensity-based measurements can be obtained for any required depth range from a few hundreds of nm to several tens of $\bm{\upmu}$m. This experimental method takes into account the effect of the spherical aberration without requiring computational correction.}

\section{Introduction}

SMLM is now a well-established method used in biology for a wide range of applications, especially single particle tracking \cite{Deich2004} in living samples and super-resolution structural observation using (f)PALM, (d)STORM or PAINT \cite{Betzig2006,Hess2006,Rust2006,vandeLinde2011,Sharonov2006}. Although retrieving the lateral positions in the focus plane is quite straightforward, 3D detection requires complementary axial information, that can be provided either by point spread function (PSF) shape measurement methods \cite{Huang2008,Pavani2009b,Shechtman2015,Juette2008,Franke2016}, in which the axial information is encoded in the shapes of the spots, or by intensity-based methods (such as interferometric measurements \cite{Shtengel2009,Aquino2011} or supercritical angle fluorescence (SAF) detection \cite{Bourg2015,Deschamps2014}), which rely on the dependence of the intensity on the depth.

In most cases, such an axial detection scheme requires a calibration to know the relationship between the measured value and the depth, or at least an experimental verification to validate the consistency of the results obtained from a theoretical characteristic curve. Most of the time, this is performed by using fluorescent beads or molecules deposited on a coverslip and scanning the objective with a piezoelectric stage to introduce defocus in the system. While inexpensive and simple to perform, this method exhibits several drawbacks arising from the refractive index mismatch between the sample and the glass coverslip.

First, the distance over which the focus plane is moved is not equal to the displacement of the objective. In practice, this so called focal shift effect produces a stretching of the apparent distances. Although theoretical formulae \cite{Sheppard1997} and experimental protocols \cite{Diaspro2002,Bratton2015} of various complexities are available to determine the value of the correction factor for different depth ranges, these methods are not sufficient to provide readily usable calibration data suitable for SMLM experiments.

Indeed, they do not account for the effect of the spherical aberration on the PSFs. Such an aberration alters the shape of the spots and thus induces a bias in the axial positions detected through PSF shaping methods. Calibrations performed by scanning the objective do not allow to record the PSFs corresponding to a realistic experimental situation where the focus plane is fixed. Several techniques have been proposed to circumvent this issue, notably by numerical computation \cite{McGorty2014}. While it does not require a cumbersome experimental procedure, it does not fully correct the effect of the spherical aberration. Deng and Shaevitz proposed a reliable experimental method using optical tweezers to axially move fluorescent beads relative to the object plane \cite{Deng2009}, at the cost of a major modification of the setup. Similarly, adaptive optics can be used to correct the spherical aberration \cite{Izeddin2012b}, but this requires expensive devices and induces a loss of photons.

\section{Method and setup}

\begin{figure}[p]
\centering
\includegraphics[height=\dimexpr \textheight - 11\baselineskip\relax]{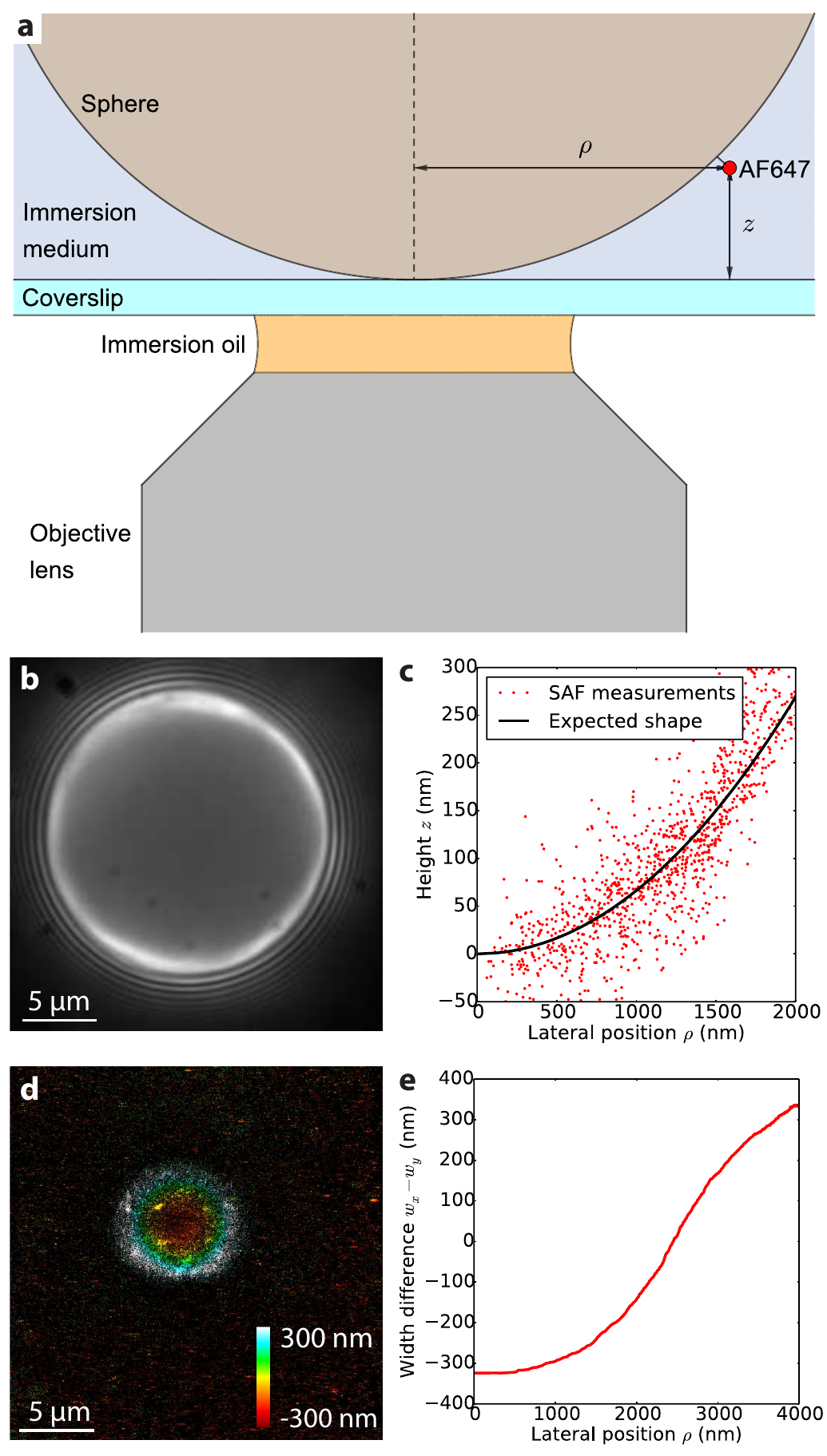}
\caption{Principle of the calibration protocol. \textbf{(a)} Schematic of the sample of coated microspheres deposited on a coverslip. \textbf{(b)} Raw diffraction limited image of a microsphere acquired in its equatorial plane (used to determine the radius and the center of the microsphere). \textbf{(c)} Depth profile detected with DONALD. The red circles correspond to the experimental data, and the black line stands for the curvature expected for a sphere of the measured radius. \textbf{(d)} Color-coded map of the width difference $\Delta w=w_x-w_y$ of the PSFs induced by the astigmatism. Note that the radius appears lower than in \textbf{(c)} since the depth of imaging limits the observation to a small portion of the sphere near the contact point with the coverslip. \textbf{(e)} Width difference $\Delta w$ profile as a function of the lateral position obtained from the sphere displayed in \textbf{(d)}.}
\label{fig1}
\end{figure}

We designed a fully experimental, sample-based calibration method to provide unbiased calibration results that can be used for 3D SMLM measurements. In this protocol, all the axial information needed is provided by the known geometry of the sample, and the acquisitions are performed in the nominal conditions, i.e. for a given objective at a given position and for given sample refractive index and fluorophore emission wavelength.

More specifically, we used 15.0 $\upmu$m ($\pm$ 1.5 $\upmu$m) diameter latex microspheres coated with biotin (Kisker Biotech, PC-BX-15.0), on which we attached the fluorophores of interest, namely Alexa Fluor (AF) 647 functionalized with streptavidin (Life Technologies, S21374). The sample was elaborated as follows: we prepared a solution containing 500 $\upmu$L of water, 500 $\upmu$L of PBS, 100 $\upmu$L of microspheres solution and 0.5 $\upmu$L of streptavidin-functionalized AF647. This solution was centrifuged 20 minutes at 13.4 krpm. The liquid was then removed and replaced with 100 $\upmu$L of PBS. By vortexing the aliquot, the deposit was dissolved. We then took 50 $\upmu$L of the final solution, delicately laid it on a glass coverslip and waited for 20 minutes after covering it to protect it from light and evaporation. Finally, we gently added 500 $\upmu$L of imaging dSTORM buffer (dSTORM smart kit, Abbelight). A similar sample was used in \cite{Bratton2015}, but only in confocal microscopy to measure the focal shift. The interest of such a sample relies on the fact that, once the lateral position of the center ($x_0$, $y_0$) and the radius $R$ of each microsphere are known, the depth of any molecule can be calculated from its measured lateral position ($x$, $y$) using the following equation (\textbf{Fig. \ref{fig1}a}):

\begin{equation}
\label{eq1}
\rho^2+(R-z)^2=R^2
\end{equation}

where $\rho=\sqrt{(x-x_0)^2+(y-y_0)^2}$ is the lateral radial position to the center of the microsphere. This directly gives the height $z$ of the molecule assuming that only the lower part of the microsphere is imaged (which is reasonable since $R$ is typically much higher than the depth of field of the system):

\begin{equation}
\label{eq2}
z=R-\sqrt{R^2-\rho^2}
\end{equation}

The sample was observed using a standard homemade super-localization microscope. The coverslip was illuminated with a 637 nm laser (Obis 637LX, 140 mW, Coherent). Our setup included a Nikon Eclipse Ti inverted microscope with a Nikon Perfect Focus System, and a Nikon Apo TIRF 100x NA1.49 objective. The camera was a 512x512-pixel EMCCD (Andor, iXon3), with one pixel corresponding to 100 nm in the object plane. We first acquired a diffraction-limited image of the sample at low illumination power (0.05 kW.cm\textsuperscript{-2}) in the equatorial plane of the sphere, i.e., at $R$ above the coverslip (\textbf{Fig. \ref{fig1}b}), which was used to measure both the radius R and the center ($x_0$, $y_0$). The effect of the diffraction was taken into account in this measurement.

We first performed an acquisition with the DONALD 3D super-resolution technique \cite{Bourg2015}, that provides absolute axial information based on the SAF intensity measurement. As this detection uses a characteristic curve obtained from a theoretical model, it does not require experimental calibration. Thus, it provides a reliable method to verify the sample shape. As seen on the $\rho$-$z$ profile (\textbf{Fig. \ref{fig1}c}), the microspheres exhibit the expected curvature in the probed area. Especially, they do not seem to significantly flatten out at the contact point with the coverslip.

Having verified the validity of the geometry of the sample, we carried out acquisitions to obtain the calibration curve of an astigmatism-based PSF shaping detection scheme. The aberration was created by a cylindrical lens added in the detection path with a focal length and a position calculated to optimize the localization precision \cite{Rieger2014}, giving a spacing of approximately 800 nm between the two focal lines. We then increased the laser excitation power (4 kW.cm\textsuperscript{-2}) to achieve a sufficient molecule density per frame (typically 1 molecule per 4 $\upmu$m\textsuperscript{2} per frame). The dSTORM data was acquired with 50 ms exposure time and 150 EMCCD gain and processed using a home-written Python code to extract the $x$ and $y$ positions of each spot, as well as their widths $w_x$ and $w_y$, via a gaussian fitting. Both the lateral and axial drifts were corrected using a cross-correlation algorithm, and the lateral deformation of the image induced by the astigmatism was accounted for. The corresponding width difference $\Delta w = w_x-w_y$ map and $\rho$-$\Delta w$ profile are presented in \textbf{Fig. \ref{fig1}d} and \textbf{Fig. \ref{fig1}e} respectively. Using $x$, $y$, $x_0$, $y_0$ and $R$, we determined the value of the depth $z$ of each molecule, which allowed us to plot the calibration curve ($w_x$, $w_y$) as a function of $z$. For the statistical pooling of the localization data, we analyzed a total of approximately 20.000 localization events.

As a comparison, we also conducted an acquisition on a sample of 20 nm fluorescent beads deposited on a glass coverslip (Invitrogen, F8783). The object plane was scanned through the sample by moving the objective thanks to a piezoelectric stage, and the axial positions were corrected in order to take into account the focal shift.

\section{Results and discussion}

\begin{figure}[p]
\centering
\includegraphics[height=\dimexpr \textheight - 9\baselineskip\relax]{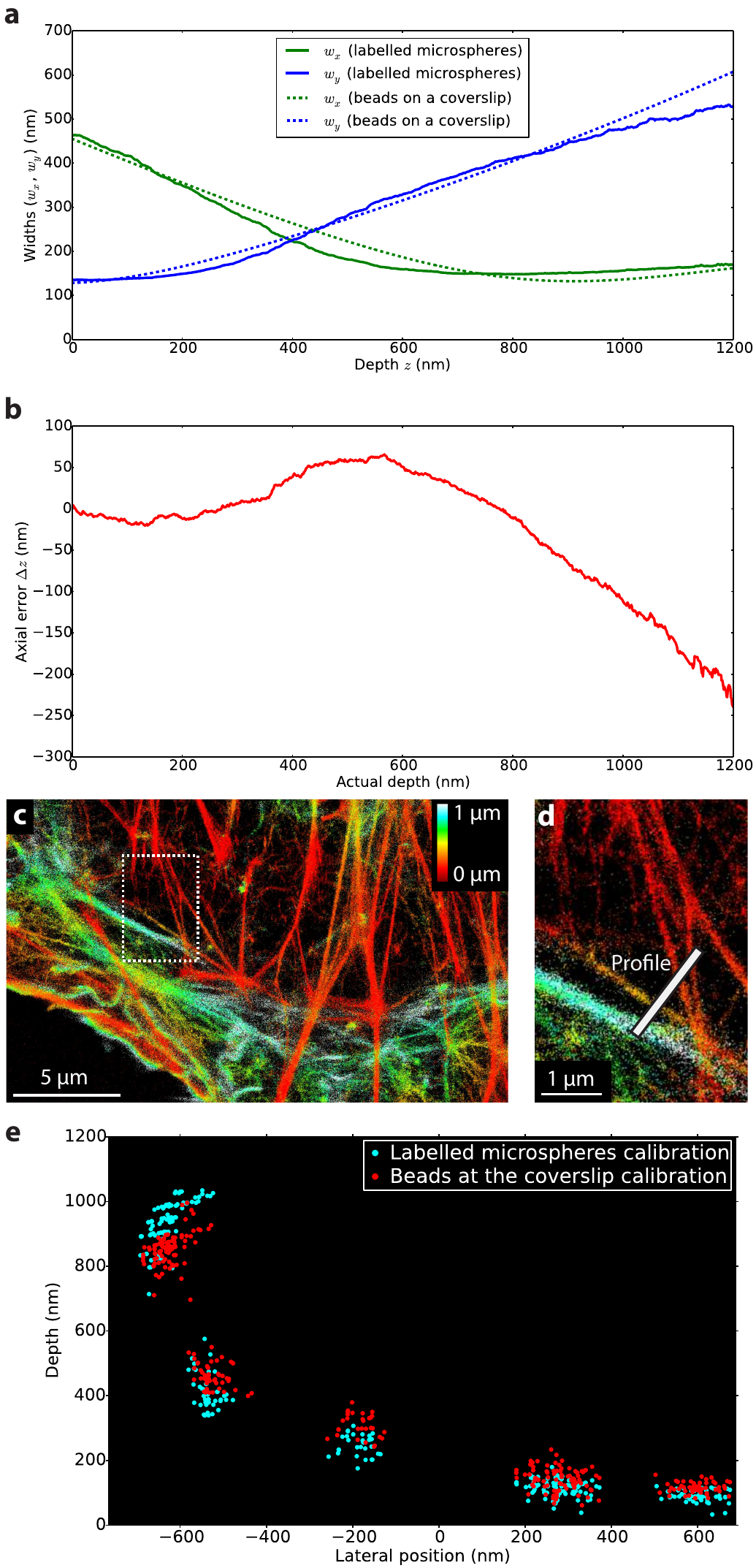}
\caption{Calibration results. \textbf{(a)} Calibration curve showing the widths $w_x$ and $w_y$ obtained by analyzing the data acquired from the labelled microspheres sample. The comparison with the method using fluorescent beads at the coverslip is also displayed. \textbf{(b)} Error made on the measured axial position when using immobilized fluorescent beads on a coverslip to perform the calibration. The focal shift induced by the index mismatch was corrected. \textbf{(c)} 3D (color-coded depth) localization image of actin labelled with AF647. \textbf{(d)} Zoom on the boxed region shown in \textbf{(c)}. \textbf{(e)} $x$-$z$ profile of actin filaments along the axis displayed in \textbf{(d)}, comparing the two calibration methods.}
\label{fig2}
\end{figure}

Thanks to the previously described method, we obtained the calibration curve of the astigmatism and compared it to the method using fluorescent beads deposited on a coverslip (\textbf{Fig. \ref{fig2}a}), which yields significantly different results. Note that because the experiment was performed in the nominal conditions, a single acquisition provides a calibration valid over the whole axial detection range of the instrument, namely 1.2 $\upmu$m. The difference between the two methods is mainly due to the influence of the spherical aberration arising from the index mismatch on the PSF shapes, which is not considered in the latter case. The resulting axial bias (\textbf{Fig. \ref{fig2}b}) proves to behave in a very nonlinear way (especially, it is either positive or negative depending on the depth), reaching values as large as 200 nm over a 1200 nm range. Such a result is in agreement with previous studies quantifying the effect of the spherical aberration in calibration experiments \cite{Deng2009,McGorty2014,Proppert2014}, both in terms of curve shape and magnitude. To illustrate the importance of this difference, we performed measurements on a COS-7 sample labelled with AF647-phalloidin to image the actin network (\textbf{Figs. \ref{fig2}c} and \textbf{\ref{fig2}d}). The slice profile presented in \textbf{Fig. \ref{fig2}e} highlights the axial bias that arises from the effect of the spherical aberration when using fluorescent beads immobilized on a coverslip to perform the calibration.

One of the key features of our technique is its precision, that makes it especially suitable for super-localization experiments. The statistical uncertainty $\sigma_z$ on the value of $z$ calculated from \textbf{eq. \ref{eq2}} decreases as $R$ increases. Indeed, the lower the local slope of the sphere surface, the lower the influence of the lateral localization precision. As it is necessary to image a whole microsphere to measure its radius and center, the optimal microsphere diameter one should choose to achieve the best precision is equal to the size of the field of view. Assuming that the center and the radius positions can be estimated with precisions around 50 nm, a 15 $\upmu$m diameter microsphere yields axial precisions $\sigma_z$ down to 15 nm over a 1.2 $\upmu$m axial range. It should be noted that since the number of localized molecules on the microsphere is around 20.000, the localization precision has a much lower influence on the value of $\sigma_z$ than any error on the measurement of the radius or the center of the sphere.

\section{Conclusion}

We have demonstrated a fully experimental unbiased method to obtain the calibration curve of any PSF shaping based axial detection scheme in 3D SMLM. While it does not require any computation, device interfacing or modification of the optical setup, it accounts for both the focal shift and the effect of the spherical aberration on the PSFs by performing acquisitions in the nominal experimental conditions on a controlled geometry sample. These features make it suitable for a broad range of uses in SMLM as it could be used in (d)STORM, (f)PALM and (DNA-)PAINT. Aside from calibration of PSF shaping methods, this technique could also be used as a means of experimental verification of the performances of techniques requiring no calibration. The method could be further improved by labelling the microspheres with several dyes to perform a sequential calibration at different emission wavelengths, thus getting rid of the chromatic aberration influence, a major bottleneck for multiple color/species 3D SMLM imaging.

\section*{Fundings}

This work has been supported by the R{\'e}gion {\^I}le-de-France in the framework of C'Nano IdF, the nanoscience competence center of Paris Region, for the NanoSAF project.

\section*{Disclosures}

N.B. and S. L.-F. are share holders in Abbelight.

\bibliographystyle{ieeetr}
\bibliography{references}

\end{document}